\documentclass[prl,twocolumn,showpacs,preprintnumbers,amsmath,amssymb]{revtex4}
\usepackage{graphicx}% Include figure files
\usepackage{dcolumn}% Align table columns on decimal point
\usepackage{bm}% bold math
\usepackage{amsthm}
\usepackage{epsfig}
\usepackage{color}
\input xy
\xyoption{all}

\newcommand{\ket}[1]{\left| #1 \right\rangle}

\newcommand{\spacer}{\rule[0cm]{0cm}{0cm}}

\newtheorem{theorem}{Theorem}

\DeclareMathOperator{\Stab}{Stab}

\newcommand{\C}{{\mathbb C}}

\newcommand{\Id}{{\rm Id}}

\newcommand{\twotwo}[4]{\left[ \begin{array}{cc} #1 & #2 \\
                        #3 & #4 \end{array} \right]}

\begin{document}

\title{Multiparty quantum states stabilized by the diagonal subgroup of
  the local unitary group}

\author{David W. Lyons}
  \email{lyons@lvc.edu}
\author{Scott N. Walck}
  \email{walck@lvc.edu}
\affiliation{Lebanon Valley College, Annville, PA 17003}

\date{21 August 2008, revised 23 September 2008}

\begin{abstract}
We classify, up to local unitary equivalence, the set of $n$-qubit
states that is stabilized by the diagonal subgroup of the local unitary
group.  We exhibit a basis for this set, parameterized by diagrams of
nonintersecting chords connecting pairs of points on a circle, and give a
criterion for when the stabilizer is precisely the diagonal subgroup and
not larger.  This investigation is part of a larger program to
partially classify entanglement type (local unitary equivalence class)
via analysis of stabilizer structure.

\end{abstract}


\maketitle

The desire to measure and classify entanglement for states of $n$-qubit
systems has been motivated by potential applications in quantum
computation and communication that utilize entanglement as a
resource~\cite{nielsenchuang,gudder03}.  More deeply, the mystery of
entanglement has played a key role in foundational questions about
quantum mechanics itself.  Because entanglement properties of
multi-qubit states are invariant under local unitary transformations,
attempts to classify entanglement lead naturally to the problem of
classifying local unitary equivalence classes of states.

The results presented in this article arise from the following
framework, utilized in~\cite{linden98,carteret00a} for 3-qubit systems
and developed further by the authors
in~\cite{minorb1,minorb2,maxstabnonprod1,maxstabnonprod2} for $n$
qubits, for approaching local unitary equivalence classification.  The
equivalence class of a state---its orbit under the local unitary group
action---is a submanifold of Hilbert space.  There is a natural
diffeomorphism 
$${\cal O}_\psi \leftrightarrow G/\Stab_\psi$$ between the orbit 
$${\cal
O}_\psi = \{g\ket{\psi}\colon g\in G\}$$ of a state $\ket{\psi}$ and the set
$G/\Stab_\psi$ of cosets of the the stabilizer subgroup 
$$\Stab_\psi = \{g\in
G\colon g\ket{\psi}=\ket{\psi}\}$$ (termed simply {\em stabilizer} hereafter) of the local unitary group
$$G=U(1)\times SU(2)^n.$$ This duality between orbits and stabilizers
provides a means of studying entanglement types (orbits) by analyzing
stabilizer subgroups.  Focusing on stabilizers affords the additional
advantage of exploiting the well-developed theory of Lie groups and
their Lie algebras of infinitesimal transformations.

In~\cite{maxstabnonprod1}, we showed that for any state $\psi$, there is
a disjoint union of the set of qubit labels
$$\{1,2,\ldots,n\}={\cal B}_1 \cup \ldots \cup {\cal B}_p \cup {\cal
  R}$$ 
so that the stabilizer (after an LU transformation, if
necessary) has the form
\begin{equation}\label{stabdecomp}
\Stab_\psi = \Delta_1 \times \cdots \times \Delta_p \times H,  
\end{equation}
where each $\Delta_j$ is a subgroup isomorphic to $SU(2)$ consisting of
elements the form 
\begin{equation}\label{diageltform}
\underbrace{1}_{\mbox{phase factor}}\times
    \underbrace{(g,\ldots,g)}_{\mbox{in qubits ${\cal B}_j$}}\times
    \underbrace{(\Id,\ldots,\Id)}_{\mbox{in qubits outside of ${\cal
    B}_j$}}  
\end{equation}
where $g$ ranges over $SU(2)$, and $H$ is a subgroup whose projection
into each $SU(2)$ factor of $G$ is trivial in qubits $\cup_j {\cal B}_j$
and is 0- or 1-dimensional in the remaining $SU(2)$ factors in qubits in
${\cal R}$\footnote{To be precise, these results are stated and proved
in terms of the stabilizer Lie subalgebra in~\cite{maxstabnonprod1}.  On
the group level, the stabilizer may also have a discrete (0-dimensional
and therefore finite, since $G$ is compact) factor.  We may ignore this
technicality for the purposes of the present discussion.}.

In previous work~\cite{maxstabnonprod1,maxstabnonprod2}, we have studied
stabilizers with maximum possible dimension in the factor $H$ in the
decomposition~(\ref{stabdecomp}) and have
shown that states with such a stabilizer have important entanglement
properties.  This leads naturally to the question of what states have
stabilizer 
$$\Delta = \{(1,g,g,\ldots,g)\colon g\in SU(2)\}$$
and what interesting entanglement properties do they have?  Evidence
that this investigation will be fruitful is the 4-qubit state
\begin{align*}
\ket{M_4} &= \frac{1}{\sqrt{6}} [
  \ket{0011} + \ket{1100} + \omega (\ket{1010} + \ket{0101}) \\
  & \hspace{2cm}    + \omega^2 (\ket{1001} + \ket{0110}) ] ,
\end{align*}
where $\omega = \exp(2 \pi i/3)$.  The state $M_4$ has stabilizer
$\Delta$ and has been shown~\cite{higuchi00,brierley07} to maximize
average two-qubit bipartite entanglement, averaged over all partitions
into 2-qubit subsystems.  With this background, we pose the problem
considered in this article.

\medskip

{\em {\bf Problem}.  Classify, up to local
unitary equivalence, the space of states whose stabilizer contains
$\Delta$.  Among these states, which have stabilizer precisely equal to
$\Delta$ and not larger?}

\medskip

Let $V_\Delta$ denote the space of states whose stabilizer contains
$\Delta$.  In the language of representation theory, $V_\Delta$ is the
trivial subrepresentation Hilbert space $(\C^2)^{\otimes n}$ under the
action of $\Delta\cong SU(2)$.  
In physical terms, regarding
qubits as spin-$1/2$ particles, $V_\Delta$ is the space of states with
zero total angular momentum\footnote{$V_\Delta$ is the zero set of the
  angular momentum operator
$$
J^2 = \left[ \frac{\hbar}{2} \sum_{i=1}^n (\sigma_x)_i \right]^2
   + \left[ \frac{\hbar}{2} \sum_{i=1}^n (\sigma_y)_i \right]^2
   + \left[ \frac{\hbar}{2} \sum_{i=1}^n (\sigma_z)_i \right]^2
$$
where $(\sigma_a)_j$ is the Pauli matrix $\sigma_a$ ($a=x$, $y$ or $z$)
acting on the $j$th qubit.  The operators $i\sum (\sigma_x)_i$, $i\sum
(\sigma_y)_i$, $i\sum (\sigma_z)_i$ form a basis for the Lie algebra of $\Delta$.}.
It is known~\cite{livineterno05} that $V_\Delta=0$
for odd $n$ and that the dimension of $V_\Delta$ for $n=2m$ is the $m$th
Catalan number
$$\dim V_\Delta = \frac{1}{m+1}{2m \choose m}.$$ It is clear that any
product of $m$ singlet states $\ket{01}-\ket{10}$ (in any pairs of
qubits) is in $V_\Delta$.  In fact, as we show below, it turns out that
all states in $V_\Delta$ are linear combinations of such states.  

\begin{figure}[h]
\begin{center}
\begin{picture}(0,0)%
\includegraphics{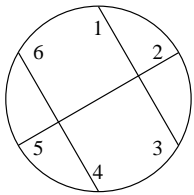}
\end{picture}%
\setlength{\unitlength}{3947sp}%
\begingroup\makeatletter\ifx\SetFigFontNFSS\undefined%
\gdef\SetFigFontNFSS#1#2#3#4#5{%
  \reset@font\fontsize{#1}{#2pt}%
  \fontfamily{#3}\fontseries{#4}\fontshape{#5}%
  \selectfont}%
\fi\endgroup%
\begin{picture}(1180,1185)(2638,-1728)
\end{picture}%

\end{center}  
\caption{Diagram for product of 3 singlet pairs specified by chords.}
\label{chorddeffig}
\end{figure}

We can represent any product of $m$ singlets by a diagram consisting of
$2m$ consecutively labeled points, joined in pairs by chords, no two of
which share an endpoint.  For example, the 6-qubit state
\begin{eqnarray*}
\ket{\psi}&=&\ket{001011}-\ket{001110}-\ket{011001}-\ket{100011}\\
&&+\ket{110001} + \ket{100110} + \ket{011100} -\ket{110100}  
\end{eqnarray*}
which is the product of singlets in qubit pairs
$\{1,3\},\{2,5\},\{4,6\}$ is shown in Figure~\ref{chorddeffig}. 

\vspace*{.2in}

\begin{figure}[h]
  \begin{center}
\begin{picture}(0,0)%
\includegraphics{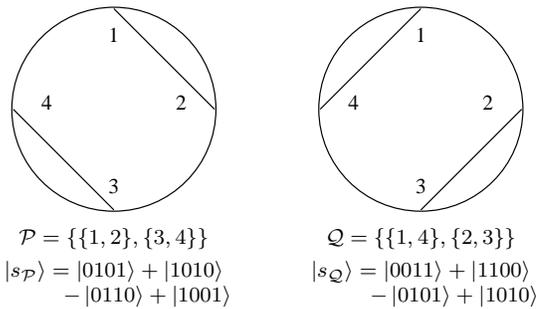}
\end{picture}%
\setlength{\unitlength}{3947sp}%
\begingroup\makeatletter\ifx\SetFigFontNFSS\undefined%
\gdef\SetFigFontNFSS#1#2#3#4#5{%
  \reset@font\fontsize{#1}{#2pt}%
  \fontfamily{#3}\fontseries{#4}\fontshape{#5}%
  \selectfont}%
\fi\endgroup%
\begin{picture}(3226,1920)(1395,-2242)
\put(2044,-2025){\makebox(0,0)[b]{\smash{{\SetFigFontNFSS{8}{9.6}{\rmdefault}{\mddefault}{\updefault}{\color[rgb]{0,0,0}$\ket{s_{\cal P}} = \ket{0101} + \ket{1010}$}%
}}}}
\put(2255,-2183){\makebox(0,0)[b]{\smash{{\SetFigFontNFSS{8}{9.6}{\rmdefault}{\mddefault}{\updefault}{\color[rgb]{0,0,0}$-\ket{0110} + \ket{1001}$}%
}}}}
\put(3972,-1814){\makebox(0,0)[b]{\smash{{\SetFigFontNFSS{8}{9.6}{\rmdefault}{\mddefault}{\updefault}{\color[rgb]{0,0,0}${\cal Q}=\{\{1,4\},\{2,3\}\}$}%
}}}}
\put(3972,-2025){\makebox(0,0)[b]{\smash{{\SetFigFontNFSS{8}{9.6}{\rmdefault}{\mddefault}{\updefault}{\color[rgb]{0,0,0}$\ket{s_{\cal Q}} = \ket{0011} + \ket{1100}$}%
}}}}
\put(4183,-2183){\makebox(0,0)[b]{\smash{{\SetFigFontNFSS{8}{9.6}{\rmdefault}{\mddefault}{\updefault}{\color[rgb]{0,0,0}$-\ket{0101} + \ket{1010}$}%
}}}}
\put(2044,-1814){\makebox(0,0)[b]{\smash{{\SetFigFontNFSS{8}{9.6}{\rmdefault}{\mddefault}{\updefault}{\color[rgb]{0,0,0}${\cal P}=\{\{1,2\},\{3,4\}\}$}%
}}}}
\end{picture}%

  \end{center}
\caption{The two nonintersecting 4-qubit chord diagrams and their associated
  singlet product states.}
\label{spillustrationfig}
\end{figure}

 Given 
a partition ${\cal P}$ of $\{1,2,\ldots,2m\}$ into 2-element subsets,
let $\ket{s_{\cal P}}$ denote the singlet product state with singlet qubit
pairs determined by ${\cal P}$.  We shall say that ${\cal P}$ {\em has no
  intersections} if the associated chord diagram has no intersecting
chords. Figure~\ref{spillustrationfig} illustrates all such states for
$m=2$. 

\medskip

Now we can state the solution to the above Problem.  Statement~1 in the
Theorem below gives a unique way to write any state whose stabilizer
contains $\Delta$, and Statement~3 asserts this representation is unique
in its local unitary equivalence class.  Statement~2 answers the second
question in the Problem above by giving a simple geometric criterion for
when a state has its stabilizer precisely equal to $\Delta$.
\begin{theorem}\label{mainresult}
\spacer
\begin{enumerate}
\item \label{spbasis} The set $\{\ket{s_{\cal P}}\colon {\cal P} \mbox{
  has no intersections}\}$ is a basis for $V_\Delta$. 
\item \label{stabnottoobig} For 
$$\ket{\psi}=\sum_{\cal P } c_{\cal P}\ket{s_{\cal P}}$$ 
for which $c_{\cal P}=0$ if ${\cal P}$ has intersections, we have
$\Stab_\psi=\Delta$ if and only if the following condition holds.
\begin{itemize}
\item [($\ast$)] For every proper subset ${\cal S}\subset
\{1,2,\ldots,2m\}$, there exists a partition ${\cal P}$ with $c_{\cal
P}\neq 0$ and some $\{a,b\}\in {\cal P}$ with $a\in {\cal S}$ and
$b\not\in{\cal S}$.
\end{itemize}
\item \label{luuniqueness} Two states $\psi,\psi'$ that are local
  unitary equivalent with $\Stab_\psi= \Stab_{\psi'} = \Delta$ are in
  fact equal up to a phase factor.
\end{enumerate}
\end{theorem}

To prove Theorem~\ref{mainresult}, we begin with a device for assigning
a particular bit string to a partition ${\cal P}$ that has no
intersections.  Given a partition ${\cal P}$, we define $I_{\cal P}$ to
the {\em smallest} (as a binary number) multi-index that occurs in the
expansion of $\ket{s_{\cal P}}$ with nonzero coefficient in the computational
basis. More generally, given an ordering ${\cal
O}=(k_1,k_2,\ldots,k_{2m})$ of the qubit labels $\{1,2,\ldots,2m\}$, we
define $I^{\cal O}_{\cal P}$ to be the smallest binary number
$i_{k_1}i_{k_2}\ldots i_{k_{2m}}$, where $I=(i_1,i_2,\ldots,i_{2m})$
ranges over the multi-indices that occur with nonzero coefficient in the
expansion of $\ket{s_{\cal P}}$ in the computational basis.  It is easy to see
how to construct $I_{\cal P}$.  For each $\{a,b\}\in {\cal P}$ with
$a<b$, assign $i_a=0$ and $i_b=1$.  Similarly, to construct $I^{\cal
O}_{\cal P}$, for each $\{k_a,k_b\}\in {\cal P}$ with $a<b$, assign
$i_{k_a}=0$ and $i_{k_b}=1$.  Observe that if ${\cal P}\neq {\cal P}'$, then
$I_{\cal P}\neq I_{{\cal P}'}$ and $I^{\cal O}_{\cal P}\neq I^{\cal
O}_{{\cal P}'}$. 
%% It is useful to consider a direction on the chords of
%% ${\cal P}$.  For $\{a,b\}\in {\cal P}$ with $a<b$, we say $a$ is the
%% {\em initial} node and $b$ is the {\em terminal} node of the chord
%% $\{a,b\}$.  Then $I_{\cal P}$ is the string that arises by assigning 0
%% to each initial node and 1 to each terminal node.
%% Figure~\ref{smalleststringfig} shows an example.

%% \begin{figure}[h]
%% \begin{center}
%% \begin{picture}(0,0)%
%% \includegraphics{figure3.ps}
%% \end{picture}%
%% \setlength{\unitlength}{3947sp}%
%% %
%% \begingroup\makeatletter\ifx\SetFigFontNFSS\undefined%
%% \gdef\SetFigFontNFSS#1#2#3#4#5{%
%%   \reset@font\fontsize{#1}{#2pt}%
%%   \fontfamily{#3}\fontseries{#4}\fontshape{#5}%
%%   \selectfont}%
%% \fi\endgroup%
%% \begin{picture}(1403,1632)(4521,-1576)
%% \end{picture}%

%% \end{center}  
%% \caption{Directed chord diagram and associated smallest string.}
%% \label{smalleststringfig}
%% \end{figure}

Now we proceed with the proof of Theorem~\ref{mainresult}.

\medskip

{\em Proof of Statement~\ref{spbasis}.}  Since the cardinality of
$\{\ket{s_{\cal P}}\colon {\cal P} \mbox{ has no intersections}\}$ is the
dimension of
$V_\Delta$~\cite{stanleyenumcombvol2,sloaneintegerseqA000108}, it
suffices to show that the $\ket{s_{\cal P}}$ are independent. 

Suppose there is a linear relation $\sum_{\cal P}c_{\cal P}\ket{s_{\cal P}}=0$
with one or more $c_{\cal P}$ nonzero.  Then there is some partition
${\cal P}_0$ with $c_{{\cal P}_0}\neq 0$ whose associated smallest
multi-index $I_{{\cal P}_0}$ is smaller than the associated multi-index
for all other partitions with $c_{\cal P}\neq 0$. The expansion of
$\sum_{\cal P}c_{\cal P}\ket{s_{\cal P}}$ in the computational basis contains
the term $\ket{I_{{\cal P}_0}}$ with nonzero coefficient, so $c_{{\cal
P}_0}$ must be zero.  This contradiction implies that there is no linear
relation $\sum_{\cal P}c_{\cal P}\ket{s_{\cal P}}=0$ with nonzero
coefficients, and independence is established. \nopagebreak
\spacer\hfill $\square$

\medskip

{\em Proof of Statement~\ref{stabnottoobig}.}  
If condition ($\ast$) in Statement~\ref{stabnottoobig} does not hold,
  then there is a set of qubits ${\cal K}$ so that every $\ket{s_{\cal P}}$
  occurring in $\ket{\psi}$ is a product of singlets in ${\cal K}$ times
  a product of singlets in the complementary set of qubits
  $\overline{{\cal K}}$.  It follows that $\Stab_\psi$ contains a product
  $\Delta_1\times \Delta_2$ of diagonal subgroups in qubits ${\cal
    K},\overline{\cal K}$ that properly contains $\Delta$.

Conversely, suppose that condition ($\ast$) holds.  Since the projection
of $\Stab_\psi$ in each $SU(2)$ factor of $G$ is 3-dimensional, we know
that $H$ in~(\ref{stabdecomp}) is trivial and therefore $\Stab_\psi$ is
a product $\Delta_1\times \cdots \times \Delta_p$ for some
$p\geq 1$~\cite{maxstabnonprod1}.  It is our aim to show that in fact,
$p=1$.  Suppose on the contrary that $p>1$, and let ${\cal K}$ be the
proper subset of $\{1,2,\ldots,2m\}$ consisting of qubits in
which $\Delta_1$ has nontrivial coordinates.  Consider the element
$$X=(0,X_1,X_2,\ldots,X_n)$$
in the Lie algebra $K_\psi$ of $\Stab_\psi$, where
$X_k=\twotwo{i}{0}{0}{-i}$ for qubits $k\in {\cal K}$ and
$X_k=0$ for qubits $k\not\in {\cal K}$.  Given a
multi-index $I=(i_1,i_2,\ldots,i_{2m})$, let
\begin{equation}\label{coeffXaction}
\alpha_I = \#\{i_k\colon i_k=0\}_ {k\in {\cal K}} - 
\#\{i_k\colon i_k=1\}_{k\in {\cal K}}.
\end{equation}
The action of  $X$ on the computational basis vector $\ket{I}$ is the following~\cite{minorb1}.
\begin{equation}\label{Xaction}
X\ket{I} = i\alpha_I\ket{I}
\end{equation}

Let ${\cal O}$ be an ordering $(k_1,k_2,\ldots,k_{2m})$ of the qubits
$\{1,2,\ldots,2m\}$ obtained by choosing $(k_1,\ldots,k_{|{\cal K}|})$ to
be any ordering of the qubits in ${\cal K}$, and choosing any ordering
$(k_{|{\cal K}|+1},\ldots,k_{2m})$ of the qubits in $\overline{\cal K}$.
Condition ($\ast$) implies that there exist one or more partitions
${\cal P}$ with $c_{\cal P}\neq 0$ and with ${\cal P}$ having at least
one chord with one end in ${\cal K}$ and the other end in
$\overline{\cal K}$.  Let $I^{\cal O}_{\cal P}$ be the associated
smallest string.  The number of 0s in $I^{\cal O}_{\cal P}$ in qubits
${\cal K}$ is the number of chords with initial ends in ${\cal K}$ and the
number of 1s in $I^{\cal O}_{\cal P}$ in ${\cal K}$ is the number of
chords with terminal ends in ${\cal K}$.  The only way to have the
number of 0s equal the number of 1s is to have all the chords that begin
in ${\cal K}$ also end in ${\cal K}$.  But the choice of ${\cal P}$
guarantees that this is not the case.  By~(\ref{Xaction}), $X$ kills a
basis vector $I$ if and only if $\alpha_I=0$ in~(\ref{coeffXaction}), so
it follows that $X\ket{I^{\cal O}_{\cal P}}\neq 0$.  Let $A$ be the set
of partitions that have at least one chord with one end in ${\cal K}$
and the other end in $\overline{\cal K}$, and let ${\cal P}_0\in A$ be
the partition whose associated smallest string $I^{\cal O}_{{\cal P}_0}$
is the smallest among all $I^{\cal O}_{\cal P}$ for ${\cal P}$ in $A$.
Since
$$
  X\ket{\psi} = X\left(\sum_{{\cal P}\in A} c_{\cal P}\ket{s_{\cal P}}\right)
$$ it follows that the term $X\ket{I^{\cal O}_{{\cal P}_0}}\neq 0$
survives in the expansion of $X\ket{\psi}$ in the computational basis.
But this contradicts the assumption that $X\in K_\psi$.  We conclude
that there must be only one $\Delta_j$ factor in $\Stab_\psi$, and we
are done.  \nopagebreak\spacer\hfill $\square$

\medskip

{\em Proof of Statement~\ref{luuniqueness}.}  
Let $g=(e^{it},g_1,\ldots,g_n)$ be a local unitary operator
such that $\ket{\psi'} = g\ket{\psi}$. The hypotheses imply that $g
\Stab_\psi = \Stab_\psi g$ or $g \Stab_\psi g^{-1} = \Stab_\psi$.  Therefore we have
$$
  g_i h g_i^{-1} = g_j h g_j^{-1}   
$$
for all $i,j$, $1\leq i,j \leq n$ and all $h\in SU(2)$,
so $g_i^{-1} g_j$ stabilizes $h$ (with respect to the action of $SU(2)$
on itself by conjugation) for all $h\in SU(2)$.  Therefore
$g_i^{-1} g_j$ must be plus or minus the identity.
The same holds for all pairs $g_i, g_j$, and so we have
\begin{eqnarray*}
  \ket{\psi'} &=& (\pm e^{i t},g_1,g_1,\ldots,g_1) \ket{\psi}\\
 &=& \pm e^{i t} \ket{\psi} 
\end{eqnarray*}
as claimed.  
\nopagebreak\spacer\hfill $\square$

\medskip

{\em Conclusion.}  We have described, in terms of a natural basis of
combinatorial objects, those states whose stabilizers contain the
diagonal subgroup of the local unitary group and have shown that
expansions in this basis are unique (up to phase) representatives of
their local unitary equivalence class.  We have also given a simple
geometric condition for when a state written in terms of this basis has
its stabilizer subgroup precisely equal to the diagonal subgroup and not
larger.  Together with previous work, the results of this paper classify
local unitary equivalence classes for states whose stabilizers are
special cases of the general stabilizer
decomposition~(\ref{stabdecomp}).  Natural next steps in this analysis
are to classify subgroups of $H$ in~(\ref{stabdecomp}) and classify the
corresponding states that have those subgroups as stabilizers, and to
classify states whose stabilizers are products of two or more factors
in~(\ref{stabdecomp}).

\medskip

{\em Acknowledgments.}  
This work has been supported by National Science Foundation grant
\#PHY-0555506 and by Lebanon Valley College Faculty Research Grants.
The authors thank the anonymous referee for numerous helpful suggestions.
The first author wishes to express thanks for conversations with
Noah Linden, Sandu Popescu and Tony Sudbery.

%%% To make a version that includes the bibliography in a single file:
%%% 1) latex filename.tex
%%% 2) cp filename.tex filenamebib.tex
%%% 3) edit filenamebib.tex to include filename.bbl after this comment
%%% 4) edit filename.tex to comment out \bibliography{bibfilename}

%\bibliography{qidwl}

\end{document}